Targeted proteomics

Pierre Lescuyer, Mireille Chevallet, Sylvie Luche, Thierry Rabilloud

DRDC/BECP, CEA grenoble, 17 rue des martyrs, 38054 GRENOBLE CEDEX 9 FRANCE
Phone: +33 438 783 212
email: thierry.rabilloud@cea.fr

The unit describes the strategy used for targeted proteomics. It relies heavily on previously published units dealing with organelle preparation, protein solubilization and proteomics techniques. A specific commentary for organelle proteomics is provided. Specific protocols for the isolation of nuclei are also provided.

Organelle proteomics

Introduction

With the increasing development of proteomics studies, it becomes more and more obvious that our current proteomics technologies are , by far, not able to cope with the enormous complexity of biological samples. There are therefore two lines of improvement. The first one is to increase the resolving power of the proteomics technologies. While this is in some sense the ideal way, it is also the most difficult and the one for which the results are most difficult to anticipate. the second way is to reduce the complexity of the sample to be analyzed, so that it gets closer to the current performances of the proteomics toolbox. For eukaryotic cells, this relies mainly on the compartimentalization of the cells into various structures such as organelles (i.e. membrane bound subcellular structures) or other components (e.g. nuclear pores, nucleoles etc…), each structure being by far less complex than the complete cell. Thus, a current trend in proteomics is to restrict the analysis to selected subcellular components rather than to complete cells. In addition to providing a better analysis depth by adjusting the complexity of the sample to the performances of the methods, this approach also provides biologically valuable informations such as subcellular localization for proteins.

Targeted proteomics relies on a series of steps, in which the first and obvious ones lead the preparation of the subcellular structure(s) with sufficient purity and reproducibility. This topic has been dealt with in detail in units 4.3 and 22.5 and will therefore not be repeated here. As the cited units deal rather with the isolation of membrane-derived organelles, protocols will be provided for isolation of nuclei.
The subsequent steps in targeted proteomics deal with the proteomic analysis per se. This has also been amply detailed in unit 22 (especially units 22.1, 22.2 and 22.4) , with specific aspects also described in units 10 (Electrophoresis) and 16 (mass spectrometry). This will be consequently not re-detailed here. However, a synthetic background discussion will deal with the specific aspects of organelle proteomics compared to the more general aspects discussed in the units mentioned above.

Strategic planning.

In the case of targeted  proteomics, the strategic planning will depend whether a single or several components isolated from the biological source is/are to be studied. For a single component, specific isolation procedures can be used which may result in damages or gross contamination of other structures. This cannot be tolerated when several components are to be analyzed. However, one component only-oriented procedures usually result in a higher purity for the target component (see below the example of nuclei).

Basic protocol

1) Isolate the component(s) of interest from the biological source. For all organelles except nuclei, follow the protocols described in units 4.3 and 22.5. For the direct isolation of nuclei, follow alternate protocol 1 or 2. For the simultaneous isolation of nuclei and other organelles, follow alternate protocol 3.

The basic rationale for the isolation of subcellular components is first to lyse the cell under conditions which do not damage the components of interest. This is why nuclei, which are limited by a lamina, can be isolated with non ionic detergents (e.g. alternate protocols 1 and 2), while membrane-limited organelles cannot. This means in turn that the use of detergents in nuclei isolation will dramatically lower the extent of contamination by organelles such as lysosomes or mitochondria. Once the cell has been opened, the components of interest are purified generally by differential centrifugation, using differences in the sedimentation velocity and/or buoyant density of the various subcellular compoents. It must be stressed, however, that there is often a significant overlap in these parameters for various subcellular components. As a result, significant cross-contamination is often experienced, for example between mitochondria and lysosomes. It is also often difficult to separate the total microsomes into the plasma membrane, reticulum and Golgi components. For such delicate final purification, an often interesting track to follow is the use of free-flow electrophoresis (Unit 22.5). This requires however a special apparatus. Furthermore, the price to pay for a high purity preparation is a low yield, and therefore an important sample consumption.

Last but not least, many subcellular components are, one way or another, linked to the cytoskeleton. As this latter component is very strongly assembled, contamination of subcellular structures with cytoskeleton is commonplace and very difficult to remove.

2) Extract the proteins as described in unit 22.4

3) Analyze the proteins by the required proteomics method, e.g. those described in units 22.1 and 22.2. the choice of the method will depend on several, often mutually exclusive, considerations, As an example, membrane proteins are poorly analyzed by techniques based on two-dimensional electrophoresis. However, alternative techniques which perform adequately with membrane proteins, such as those based on SDS electrophoresis followed by LC/MS-MS, are poorly quantitative and offer poor performance in the analysis of post-translational modifications. The final choice of the experimental proteomics approach will then depend on the aspect that is judged as most important in the study to be carried out.

Alternate protocol 1: direct isolation of nuclei from cultured cells

This protocol is intended for the rapid and direct isolation of nuclei from cultured cells. It is based on the use of nonionic detergents such as Triton X100 to dissolve the lipid bilayers present in the cells, including of course the plasma membrane but also all the inner membrane-bound inner organelles (e.g. Golgi, reticulum, mitochondria and so on) which are thus irreversibly lost. It also dissolves the reticulum-derived cisternae forming the outer nuclear membrane, usually resulting in pure nuclei.

Materials and solutions list

cell washing solution.
cell lysis solution
nuclei wash solution
nuclei storage solution

Protocol.

Harvest the cells. Collect by centrifugation at 1000g for 5 minutes at 4°C.  Wash them twice in ice-cold cell washing solution by resuspension and recentrifugation as above. Estimate the volume of the packed cell pellet.

Resuspend the cell pellet in 10 times its volume of ice-cold cell lysis solution. Incubate on ice for 20 minutes with vortexing for 20 seconds every 5 minutes. Collect the nuclei by centrifugation at 1000g for 5 minutes at 4°C. Resuspend once in an identical volume of ice-cold cell lysis solution. Incubate on ice for 5 mionutes and recentrifuge as above. resuspend in an identical volume of ice-cold nuclei wash solution, let stand on ice for 5 minutes and recentrifuge as above. Repeat this rinse once.
After the last centrifugation, estimate the nuclei pellet volume and resuspend thoroughly in the same volume of storage solution. Dispense in aliquots and store at -80°c, or better in liquid nitrogen.

Alternate protocol 2: direct isolation of nuclei from tissues solutions: same as in alternate protocol 1.

protocol:
cut the tissue of interest in small pieces with scissors. Wash the pieces with cell washing solution. Homogeneize the  tissue pieces in 10 times their columes of cell lysis solution with the help of a Dounce glass-glass homogeneizer . Ten strokes are usually sufficient. Collect and wash the nuclei as described in the basic protocol

Alternate protocol 3: cleaning nuclei isolated by other procedures.

the basic protocol and alternate protocol 1 use  direct disruption of the cells in a buffer containing Triton X100. As this results in the irreversible loss of most other organelles, these protocols cannot be used when other organelles than the nuclei are also to be collected. In this case, cell disruption protocols respecting the integrity of the organelles are used (see unit 4.3). These protocols generally provide a nuclei-rich pellet at the end of the first low speed centrifugation. This pellet contains nuclei, unbroken cells and contaminating organelles (e.g. mitochondria or membrane sheets). Cleaning of the nuclei is therefore advisable.

Solutions: same as in alternate protocol 1.

protocol: resuspend the nuclei pellet in 10 times its volume with the help of a Dounce glass-glass homogeneizer . Five strokes are usually sufficient. Collect and wash the nuclei as described in the basic protocol

Support protocol: evaluation of nuclei purity.
because of the use of a detergent, the nuclei are generally devoid of contamination by other organelles. The specific enzymes test described in unit 4.3 does not provide an accurate estimate for purity. The most likely contaminants for these nuclei preparation are cytoskeletal proteins and ribosomes. Although these are difficult to quantitate, a good idea of nuclei purity can be provided by microscopy examination using the following protocol.

Add one drop of microscopy stain solution to one drop of nuclei suspension on a microscope slide. Cover with a coverslip and leave at room temperature for 10 minutes before examination on a microsope. Magnification should be at least 200x, and phase contrast affords additional information.

Nuclei appear in light green. One or two pink-colored nucleolae should be visible inside the nuclei. Contaminatiing cytoplasm results in pink material (i.e. ribosome-containing) outside the nuclei. Cytoskeleton is visible under phase contrast as refringent filaments.

Reagents and solutions:

cell washing solution: 10 mM phosphate buffer pH 7.5, 0.125M NaCl cell lysis solution: HEPES-NaOH 10 mM pH 7.5, dithiothreitol 1mM, spermidine (trihydrochloride) 1mM, spermine (tetrahydrochloride) 0.25mM, EDTA 0.5 mM, Triton X 100 0.1% (w/v). Add 30$\mu$M of 3,4 dichloroisocoumarin (Fluka, predissolved at 30mM in dimethylformamide and kept as stock at -20°C) just before use.

nuclei wash solution: HEPES-NaOH 10 mM pH 7.5, dithiothreitol 1mM, sucrose 0.2M, MgCl2 2mM.

nuclei storage solution: HEPES-NaOH 10 mM pH 7.5, dithiothreitol 1mM, glycerol 25% (v/v), MgCl2 5mM

microscopy stain solution: dissolve 150 mg of methyl green and 250 mg of pyronin Y in 2.5 ml of 95% ethanol. Add 50mM acetate buffer pH 4.8 to 100ml. Add then 40g of sucrose and dissolve by mixing. Keep at 4°C in tightly stoppered flasks (to avoid evaporation of the acetic acid)

Background information:

As mentioned above, the rationale in targeted proteomics is to analyze a defined subset of cellular proteins in order to adapt the complexity of the sample to the performances of the analysis tool. Although this trend appeared some time ago (Wilkins et al. 1995, Gygi et al. 2000), it becomes more and more apparent that the proteomics toolbox cannot face the complexity of the real biological samples. This holds true for all the methods described to date, whether they are based on 2D gels (Gygi et al. 2000) or not (Wu et al. 2003).
From this evidence, the idea of analyzing only cellular subsets of proteins became obvious a few years ago (Rabilloud et al. 1998, Jung et al. 2000). These subsets can be obtained from biochemical fractionation of the proteins of the complete sample (Fountoulakis et al. 1997, Zuo et al. 2002), but the performances of the classical biochemical fractionation become limiting very soon. Another approach, which is represented by organelle proteomics, takes advantage of the existing compartimentalization of the eukaryotic cells. In addition to providing samples of reduced complexity for proteomics analysis, organelle proteomics provides cell biology-relevant informations, e.g. the presence of "new" or "unexpected" proteins in a given organelle. This can give rise in turn to important observations in terms of cell biology (Gagnon et al. 2002).
The relevance of organelle proteomics depends in turn on the purity and reproducibility of the organelle preparation, and these parameters obviously play a role at different levels for modulating the quality of the final proteomics analysis.

Critical parameters

The first, obvious level at which organelle purity plays its role is of course the accuracy of the analysis. A low purity will of course results in contaminating proteins being present in the organelle map. This can lead to misinterpretation of the results, especially when unexpected proteins are present in the map.
Another important consequence of organelle purity is the depth of the resulting proteomics analysis. A good example of this phenomenon is provided by the analysis of mitochondria (Zischka et al. 2003). Even for organelle proteomics, the limiting factor is the capacity of the proteomics toolbox. Consequently, when organelle of low purity are analyzed, contaminating proteins occupy a substantial fraction of the analysis space (2D gels or peptide MS peaks) and prevent minor, bona fide organelle proteins from being analyzed. This problem gradually disappears when the organelle purity rises, allowing the analysis of minor proteins with pure organelles (Zischka et al. 2003), which is always a challenge in proteomics.

The second critical point is of course the reproducibility of the organelle preparation. While a poor reproducibility is a minor problem for cartographic proteomics purposes (i.e. what is the protein composition of a given sample), it becomes a major problem when comparative, physiology-oriented studies are carried out, as in clinical proteomics. This reproducibility problem can be really tricky to solve, as the changes induced by the pathology studied or upon the the physiological process of interest can interfere with the organelle preparation. As an example, any change involving a major change in the cytoskeleton, a major player and problem in organelle preparation, will alter the purity of the final organelle preparation. This will result in turn into changes in the proteomics map when comparing the different physiological situations of interest. However, these changes are artefactual and, although linked to a change in the cell physiology, are not really related to the organelle of interest. The robustness of the protocol used for organelle preparation is therefore quite critical in such transversal studies, and rather difficult to establish, as a protocol performing well with control samples can fail insidiously with other samples used in the study.

Another major caveat in targeted proteomics is sample consumption, as examplified by proteomics of nuclear proteins. Nuclei represent 10% of the total cell protein mass, and the isolation procedures have a yield close to 50%. This means in turn that 5% of the initial sample is used at the end, and 95% of the protein content is lost. While such losses can be dealt with in some experimental models, they are usually not tolerable when the sample supply is limited, e.g. in clinical proteomics.

These problems lead in turn to the choice of the technology to be used for targeted proteomics. This is obviously linked to the aims of the study to be performed. 2D gel-based technologies are the best choice when quantitative changes and/or changes in post translational modifications are important in the design of the study. As mentioned above, this is seldom the case in current targeted proteomics. Moreover, these technologies have usually a high sample consumption and a strong chemical bias against some classes of proteins, such as membrane proteins. Conversely, peptide separation-based approaches (e.g. in Washburn et al. 2001), do not have the same drawbacks as 2D gels and are therefore highly complementary. Their interest in targeted proteomics is linked to their low sample consumption (Pflieger et al. 2002), and to the fact that they do not have the same bias against classes of proteins. Membrane proteins present in the sample are therefore more easily analyzed (Pflieger et al. 2002). These positive features are shared both by methods relying only on peptide separation (Washburn et al. 2001), and by methods using a first step of protein separation by SDS electrophoresis and then a peptide separation step (e.g. in Rout et al. 2000, Bell et al. 2001, Pflieger et al. 2002 ), which seem easier to implement that the peptide-only approaches (Pflieger et al. 2002).

Anticipated results

Keeping in mind these caveats linked to targeted proteomics, the benefits of targeted proteomics compared to global proteomics can be questioned. However, targeted proteomics brings invaluable information which is out of reach by global proteomics methods.

The first type of information, on a structural point of view, is the identification of the components of the component of interest. Recent examples can be found in organelle proteomics on the mitochondria (Taylor et al. 2002, Lescuyer et al. 2003), on the golgi (Bell et al. 2001) or on the peroxisome (Yi et al. 2002). On the side of non-orrganelle subcellular components, examples can be found on the nuclear pore (Rout et al. 2000 or on the nucleolus (Andersen et al. 2002, Scherl et al. 2002).
These studies also point out the fact that the proteomics part is even not the most difficult part. the validation of the data, i.e. the confirmation of the assignment made to the organelle via the proteomics approach, is much more difficult and time-consuming (Rout et al. 2000). In some occasions, the proteomics approach points to unexpected but verified localizations of proteins. Such events can be the entry point to more cell physiology-oriented studies, as exemplified in Gagnon et al. (2002).

The second stage of targeted proteomics lies in more dynamic studies, in which changes in the organelle proteome occurring during various physiological conditions are investigated. Examples of this approach can be found even before the word proteomics even existed (Howe and Solter 1980, Rabilloud et al. 1991). These approaches are however difficult to use because of the sample consumption, as outlined above.

As a conclusion, these questions in the best choice of a subtype of proteomics approach for targeted proteomics must not mask the fact that the real bottleneck in the field is clearly the reproducible preparation of pure subcellular structures.